\documentclass[journal, twoside]{IEEEtran}
\usepackage{times}
\usepackage{amsmath}
\usepackage{amssymb}
\usepackage{epsfig,latexsym,subfigure,cite}
\usepackage{color}

\providecommand{\xv}{\mathbf{x}}
\providecommand{\yv}{\mathbf{y}}
\providecommand{\uv}{\mathbf{u}}
\providecommand{\vv}{\mathbf{v}}
\providecommand{\sv}{\mathbf{s}}
\providecommand{\Uv}{\mathbf{U}}
\providecommand{\Vv}{\mathbf{V}}
\providecommand{\Xv}{\mathbf{X}}
\providecommand{\Yv}{\mathbf{Y}}
\providecommand{\Sv}{\mathbf{S}}
\providecommand{\Zv}{\mathbf{Z}}
\providecommand{\Xc}{{\mathcal X}}
\providecommand{\Yc}{{\mathcal Y}}
\providecommand{\Wc}{{\mathcal W}}

\providecommand{\Cc}{{\mathcal C}}

\providecommand{\Nc}{{\mathcal N}}

\providecommand{\Ycc}{{\tilde \Yv}}
\providecommand{\Xcc}{{\tilde \Xv}}
\providecommand{\Zcc}{{\tilde \Zv}}
\providecommand{\Scc}{{\tilde \Sv}}
\providecommand{\scc}{{\tilde \sv}}
\providecommand{\Cs}{{\mathbb C}}
\providecommand{\Rs}{{\mathbb R}}

\providecommand{\A}{A^{(n)}_{\epsilon}}
\providecommand{\Pe}{P^{(n)}_{e}}

\providecommand{\E}{\normalfont\textsf{E}}

\newtheorem{theorem}{Theorem}
\newtheorem{corollary}{Corollary}
\newtheorem{lemma}{Lemma}

\newtheorem{remark}{Remark}
\newtheorem{example}{Example}

\begin{document}

\title{Discrete Memoryless Interference and Broadcast Channels with Confidential
Messages:\\
Secrecy Rate Regions}

\author{
Ruoheng~Liu,~\IEEEmembership{Member,~IEEE},
Ivana~Mari\'c,~\IEEEmembership{Member,~IEEE},
Predrag~Spasojevi\'c,~\IEEEmembership{Member,~IEEE},
and Roy~D.~Yates~\IEEEmembership{Member,~IEEE}
\thanks{Manuscript received February  16, 2007; revised July 30, 2007 and October 15, 2007.
This work was supported by NSF Grant ANI 0338805. The material in this paper was
presented in part at the 44th Allerton Conference on Communication, Control, and
Computing, Urbana, IL, USA, September, 2006 and
Information Theory and Application Workshop (ITA), San Diego, CA, USA Jan-Feb, 2007.}
\thanks{R.\ Liu is with Department of Electrical Engineering, Princeton
University, Princeton, NJ 08544 USA (email:~rliu@princeton.edu).}
\thanks{P.\ Spasojevi\'c and R.\ D.\ Yates are with WINLAB, Electrical and Computer
Engineering, Rutgers University, North Brunswick, NJ 08902 USA (email:~
\{spasojev,ryates\}@winlab.rutgers.edu).}
\thanks{I.\ Mari\'c is with WSL, Department of Electrical Engineering, Stanford University, Stanford, CA
94305 (email:~ivanam@wsl.stanford.edu).}
}

\markboth{IEEE TRANSACTIONS ON INFORMATION THEORY, VOL. X, NO. X, XXX 2008}%
{Discrete Memoryless Interference and Broadcast Channels with Confidential Messages:
Secrecy Rate Regions}

\maketitle


\begin{abstract}
We study information-theoretic security for discrete memoryless {\it
interference} and {\it broadcast} channels with independent confidential
messages sent to two receivers. Confidential messages are transmitted to their
respective receivers with {\it information-theoretic secrecy}. That is, each
receiver is kept in total ignorance with respect to the message intended for
the other receiver. The secrecy level is measured by the equivocation rate at
the eavesdropping receiver. In this paper, we present inner and outer bounds on
secrecy capacity regions for these two communication systems. The derived outer
bounds have an identical mutual information expression that applies to both
channel models. The difference is in the input distributions over which the
expression is optimized. The inner bound rate regions are achieved by {\it
random binning} techniques. For the broadcast channel, a {\it double-binning}
coding scheme allows for both joint encoding and preserving of confidentiality.
Furthermore, we show that, for a special case of the interference channel,
referred to as the {\it switch} channel, the two bound bounds meet. Finally, we
describe several transmission schemes for Gaussian interference channels and
derive their achievable rate regions while ensuring mutual
information-theoretic secrecy. An encoding scheme in which transmitters
dedicate some of their power to create {\it artificial noise} is proposed and
shown to outperform both time-sharing and simple multiplexed transmission of
the confidential messages.
\end{abstract}

\section{Introduction}

The broadcast nature of a wireless medium allows for the transmitted
signal to be received by all users within the communication range.
Hence, wireless communication sessions are very susceptible to
eavesdropping. The information-theoretic single user secure
communication problem was first characterized using the {\it wiretap
channel} model proposed by Wyner \cite{Wyner75}. In this model, a
single source-destination communication link is eavesdropped by a
wiretapper via a degraded channel. The secrecy level is measured by
the equivocation rate at the wiretapper. Wyner showed that secure
communication is possible without sharing a secret key between
legitimate users, and determined the tradeoff between the
transmission rate and the secrecy level \cite{Wyner75}. This result
was generalized by Csisz{\'{a}r and K{\"{o}rner who determined the
capacity region of the broadcast channel with confidential messages
\cite{CsiszarKorner78} in which a message intended for one of the
receivers is confidential.

Following the work of Wyner \cite{Wyner75} and Csisz{\'{a}r and K{\"{o}rner
\cite{CsiszarKorner78}, the more recent information-theoretic research on
secure communication focuses at implementing security on the physical layer.
Based on independent efforts, the authors of \cite{Liang06isit} and
\cite{Liu:ISIT:06} described achievable secure rate regions and outer bounds
for a two-user discrete memoryless multiple access channel with confidential
messages. This model generalizes the multiple access channel (MAC) \cite[Sec.
14.3]{Cover} by allowing each user (or one of the users) to receive noisy
channel outputs and, hence, to eavesdrop the confidential information sent by
the other user. In addition, the Gaussian MAC wiretap channel has been analyzed
in
\cite{Liang06it,Tekin:arxiv:06,Tekin:ISIT:06,Tekin:allerton:06,Tekin:ITA:07}.
The relay channel with confidential messages where the relay node acts as both
a helper and a wiretapper has been considered in \cite{Oohama:ITW:01}. The
relay-eavesdropper channel has been proposed in \cite{Lai06it}. More recently,
the cognitive interference channel with confidential messages has been
addressed in \cite{Liang:Allerton:2007}. The effects of fading on secure
wireless communication have been studied in
\cite{Barros:ISIT:06,Liang06allerton_a,Gopala:ISIT:07,Li:ISIT:07,Tang:ISIT:07}.

In this paper, we study two distinct but related in multi-terminal secure
communication problems following the information-theoretic approach. We focus
on discrete memoryless \emph{interference} and \emph{broadcast} channels with
independent confidential messages sent to two receivers. Confidential messages
are transmitted to their respective receivers while ensuring mutual
\emph{information-theoretic secrecy}. That is, each receiver is kept in total
ignorance with respect to the message intended for the other receiver. We first
derive outer bounds on capacity regions for these two communication systems.
These bounds have an identical mutual information expression. The expression is
optimized over different input distributions, i.e., for the interference
channel, the two senders offer independent inputs to the channel and, for the
broadcast channel, the sender jointly encodes both messages. We also derive
achievable rate regions for the two channel models. Here, we only consider
sending confidential messages and, hence, no common message in the sense of
Marton \cite{Marton79} is conveyed to the receivers in the case of the
broadcast channel. The inner bounds are achieved using {\it random binning}
techniques. For the broadcast channel, a {\it double-binning} coding scheme
which allows for both joint precoding as in the classical  broadcast channel
\cite{Marton79}, and preserving of confidentiality. Similarly, ensuring of
confidential messages precludes partial decoding of the message intended for
the other receiver in the case of the interference channel. Hence,
rate-splitting encoding used by Carleial \cite{Carleial78} and Han and
Kobayashi \cite{HanKobayashi81} employed with the classical interference
channel is precluded. Instead, the encoders will use only stochastic encoders.
We show that for the special case of the interference channel, referred to as
the {\it switch} channel, derived bounds meet. Finally, we describe several
transmission schemes for general Gaussian interference channels and derive
their achievable rate regions while still ensuring information-theoretic
secrecy. An encoding scheme in which transmitters dedicate some of their power
to create {\it artificial noise} is proposed and shown to outperform both
time-sharing and simple multiplexed transmission of the confidential messages.

The remainder of this paper is organized as follows. The notation and the
channel model are given in Sec.~\ref{sec:model}. We state the main results in
Sec.~\ref{sec:result}. Outer bounds  are derived in Sec.~\ref{sec:outer}. Inner
bounds associated with the achievable coding scheme for the interference and
broadcast channels with confidential messages are established in
Sec.~\ref{sec:inner}. Finally, the results are summarized in
Sec.~\ref{sec:con}.

\section{Definitions and Notations} \label{sec:model}

\subsection{Notations}

Throughout the paper, a random variable is denoted with an upper case letter
(e.g., $X$), its realization is denoted with the corresponding lower case
letter (e.g., $x$), the finite alphabet of the random variable is denoted with
the corresponding calligraphic letter (e.g., $\Xc$), and its probability
distribution is denoted with $P_X(x)$. For example, the random variable $X$
with probability distribution $P(x)=P_{X}(x)$ takes on values in the finite
alphabet $\Xc$. A boldface symbol denotes a sequence with the following
conventions
\begin{align*}
&  &\Xv&=[X_{1}, \dots, X_{n}],\quad \Xv^{i}=[X_{1}, \dots, X_{i}],&\\
&\text{and}& \Xcc^{i}&=[X_{i}, \dots, X_{n}].&
\end{align*}
Finally, we use $\A(P_{X})$ to denote the set of (weakly) jointly typical
sequences $\xv$ with respect to $P(x)$ (see \cite{Cover} for more details).

\subsection{The Interference Channel with Confidential Messages}

Consider a discrete memoryless interference channel with finite input alphabets
$\Xc_1$, $\Xc_2$, finite output alphabets $\Yc_1$, $\Yc_2$, and the channel
transition probability distribution $P(y_1,y_2|x_1,x_2)$. Two transmitters wish
to send independent, confidential messages to their respective receivers. We
refer to such a channel as the {\it interference channel with confidential
messages} (IC-CM). This communication model is shown in Fig.~\ref{fig:md2}.
Symbols $(x_1,x_2)\in(\Xc_1\times\Xc_2)$ are the channel inputs at transmitters
$1$ and $2$, and $(y_1,y_2)\in(\Yc_1\times\Yc_2)$ are the channel outputs at
receivers $1$ and $2$, respectively.
\begin{figure}[bt]
 \centerline{\includegraphics[width=0.95\linewidth,draft=false]{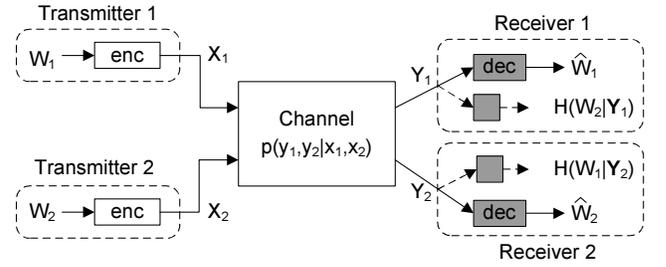}}
  \caption{Interference Channel with Confidential Messages.}
  \label{fig:md2}
\end{figure}

Transmitter $t$, $t=1,2$, intends to send an independent message
$W_t\in\{1,\dots,M_t\}$ to the desired receiver $t$ in $n$ channel uses while
ensuring information-theoretic secrecy. The channel is memoryless in the sense
that
\begin{align*}
P(\yv_1,\yv_2|\xv_1,\xv_2)=\prod_{i=1}^{n}P(y_{1,i},y_{2,i}|x_{1,i},x_{2,i}).
\end{align*}
A stochastic encoder for transmitter $t$ is described by a matrix of
conditional probabilities $f_t(\xv_t|w_t)$, where $\xv_t\in\Xc_t^n$, $w_t\in
\Wc_t$, and
\begin{align*}
\sum_{\xv_t \in \Xc_t^n}f_t(\xv_t|w_t)=1.
\end{align*}
Decoding functions are mappings $\psi_t\, :\, \Yc_t\rightarrow \Wc_t$. Secrecy
levels at receivers $1$ and $2$ are measured with respect to the equivocation
rates
\begin{equation} \label{eq:nq}
\frac{1}{n}H(W_2|\Yv_1)\quad \text{and} \quad \frac{1}{n}H(W_1|\Yv_2).
\end{equation}

An $(M_1,M_2,n,\Pe)$ code for the interference channel consists of two encoding
functions $f_1$, $f_2$, two decoding functions $\psi_1$, $\psi_2$, and the
corresponding maximum average error probability
\begin{align}
\Pe \triangleq \max \{P_{e,1}^{(n)},\,P_{e,2}^{(n)}\} \label{eq:defPe}
\end{align}
where, for $t=1,2$,
\begin{align*}
P_{e,t}^{(n)}&=\sum_{w_1,w_2}\frac{1}{M_1M_2}P\bigl[\psi_t(\Yv_t)\neq
w_t|(w_1,w_2)~\text{sent} \bigr].
\end{align*}
A rate pair $(R_1,\, R_2)$ is said to be \emph{achievable} for the interference
channel with confidential messages if, for any $\epsilon_0>0$, there exists a
$(M_1, M_2, n, \Pe)$ code such that
\begin{align}
M_t\ge 2^{nR_t}\quad \text{for}~t=1,2 \label{eq:Mr}
\end{align}
and the reliability requirement
\begin{align}
\Pe&\le \epsilon_0 \label{eq:MT}
\end{align}
and the security constraints
\begin{subequations}\label{eq:equiv}
\begin{align}
nR_1-H(W_1|\Yv_2)&\le n\epsilon_0 \label{eq:equiv1}\\
nR_2-H(W_2|\Yv_1)&\le n\epsilon_0 \label{eq:equiv2}
\end{align}
\end{subequations}
are satisfied. This definition corresponds to the so-called {\it weak secrecy-key rate}
\cite{Maurer:EUROCRYPT:00}. A stronger measurement of the secrecy level has been defined
by Maurer and Wolf in terms of the absolute equivocation \cite{Maurer:EUROCRYPT:00},
where the authors have shown that the former definition could be replaced by the latter
without any rate penalty for the wiretap channel.

The capacity region of the IC-CM is the closure of the set of all achievable
rate pairs $(R_1,R_2)$, denoted by $\Cs_{\rm IC}$.

\subsection{The Broadcast Channel}

\begin{figure}[bt]
 \centerline{\includegraphics[width=0.9\linewidth,draft=false]{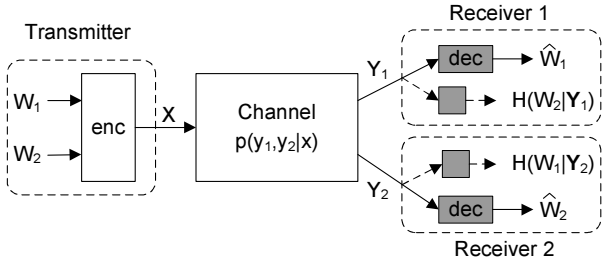}}
  \caption{Broadcast Channel with Confidential Messages.}
  \label{fig:BC}
\end{figure}

We also consider the {\it broadcast channel with confidential messages} (BC-CM)
in which secret messages from a single transmitter are to be communicated to
two receivers, as shown in Fig.~\ref{fig:BC}. A discrete memoryless BC-CM is
described using finite sets $\Xc$, $\Yc_1$, $\Yc_2,$ and a conditional
probability distribution $P(y_1,y_2|x)$. Symbols $x\in \Xc$ are channel inputs
and $(y_1,y_2)\in(\Yc_1\times\Yc_2)$ are channel outputs at receivers $1$ and
$2$, respectively. The transmitter intends to send an independent message
$W_t\in\{1,\dots,M_t\}\triangleq \Wc_t$ to the respective receiver
$t\in\{1,\,2\}$ in $n$ channel uses while ensuring information-theoretic
secrecy as given by (\ref{eq:equiv1}) and (\ref{eq:equiv2}). The channel is
memoryless in the sense that
\begin{align*}
P(\yv_1,\yv_2|\xv)=\prod_{i=1}^{n}P(y_{1,i},y_{2,i}|x_i).
\end{align*}
A stochastic encoder is specified by a matrix of conditional probabilities
$f(\xv|w_1,w_2)$, where $\xv\in\Xc^n$, $w_t\in \Wc_t$, and
\begin{align*}
\sum_{\xv \in \Xc^n}f(\xv|w_1,w_2)=1.
\end{align*}
Note that $f(\xv|w_1,w_2)$ is the probability that the pair of messages
$(w_1,w_2)$ are encoded as the channel input $\xv$. The decoding function at
the receiver $t$ is a mapping $\phi_t\, :\, \Yc_t\rightarrow \Wc_t$.

The secrecy levels of confidential messages $W_2$ and $W_1$ are measured,
respectively, at receivers 1 and 2 in terms of the equivocation rates
(\ref{eq:nq}). An $(M_1,M_2,n,\Pe)$ code for the broadcast channel consists of
the encoding function $f$, decoding functions $\phi_1$, $\phi_2$, and the
maximum error probability $P^{(n)}_{e}$ in (\ref{eq:defPe}), where, for
$t=1,2$,
\begin{align}
P^{(n)}_{e,t}&=\sum_{w_1,w_2}\frac{1}{M_1M_2}P\bigl[\phi_t(\Yv_t)\neq
w_t|(w_1,w_2)~\text{sent} \bigr].
\end{align}
A rate pair $(R_1,\, R_2)$ is said to be achievable for the broadcast channel with
confidential messages if, for any $\epsilon_0>0$, there exists a $(M_1, M_2, n, \Pe)$
code which satisfies (\ref{eq:Mr})--(\ref{eq:equiv}).

The capacity region of the BC-CM is the closure of the set of all achievable
rate pairs $(R_1,R_2)$, denoted by $\Cs_{\rm BC}$.

\section{Main Results} \label{sec:result}

In this section, we state our main results. We first describe the outer and
inner bounds on the capacity regions of both interference and broadcast
channels with confidential messages. We then show that the derived bounds meet
for a special case of the interference channel, called the switch channel.
Finally, we propose several transmission schemes for Gaussian interference
channels and derive their achievable rate regions under information-theoretic
secrecy.

\subsection{Interference Channel with Confidential Messages}

Let $U$, $V_1$, and $V_2$ be auxiliary random variables. We consider the
following two classes of joint distributions for the interference channel. Let
$\pi_{\rm IC-O}$ be the class of distributions $P(u,v_1,v_2,x_1,x_2,y_1,y_2)$
that factor as
\begin{equation} \label{eq:cIC}
P(u)P(v_1,v_2|u)P(x_1|v_1)P(x_2|v_2)P(y_1,y_2|x_1,x_2),
\end{equation}
and $\pi_{\rm IC-I}$ be the class of distributions that factor as
\begin{equation} \label{eq:cICa}
P(u)P(v_1|u)P(v_2|u)P(x_1|v_1)P(x_2|v_2)P(y_1,y_2|x_1,x_2).
\end{equation}


\begin{theorem}\label{thm:IC-O} [outer bound for IC-CM]
Let ${\Rs}_{\rm O}(\pi_{\rm IC-O})$ denote the union of all $(R_1,R_2)$ satisfying
\begin{subequations} \label{eq:ob-IC}
\begin{align}
0\le R_1&\le \min\left[
              \begin{array}{l}
                I(V_1;Y_1|U) -I(V_1;Y_2|U), \\
                I(V_1;Y_1|V_2,U) -I(V_1;Y_2|V_2,U)
              \end{array}
            \right]   \label{eq:ob-IC-a}\\
0\le R_2&\le \min\left[
              \begin{array}{l}
                I(V_2;Y_2|U)-I(V_2;Y_1| U), \\
                I(V_2;Y_2|V_1,U) -I(V_2;Y_1|V_1,U)
              \end{array} \label{eq:ob-IC-b}
            \right]
\end{align}
\end{subequations}
over all distributions $P(u,v_1,v_2,x_1,x_2,y_1,y_2)$ in $\pi_{\rm
IC-O}$. For the interference channel $(\Xc_1 \times \Xc_2, P(y_1,y_2|x_1,x_2),
\Yc_1\times \Yc_2)$ with confidential messages, the capacity region
$$\Cs_{\rm IC}\subseteq \Rs_{\rm O}(\pi_{\rm IC-O}).$$
\end{theorem}
\begin{IEEEproof}
We provide the proof of Theorem~\ref{thm:IC-O} in Sec.~\ref{sec:outer}.
\end{IEEEproof}

\begin{theorem}\label{thm:IC-I} [inner bound for IC-CM]
Let ${\Rs}_{\rm IC}(\pi_{\rm IC-I})$ denote the union of all $(R_1,R_2)$ satisfying
\begin{subequations} \label{eq:IC-IN}
\begin{align}
0 &\le R_1 \le   I(V_1;Y_1|U)-I(V_1;Y_2|V_2,U)   \label{eq:IC-IN-a}\\
0 &\le R_2 \le   I(V_2;Y_2|U)-I(V_2;Y_1|V_1,U)\label{eq:IC-IN-b}
\end{align}
\end{subequations}
over all distributions $P(u,v_1,v_2,x_1,x_2,y_1,y_2)$ in $\pi_{\rm IC-I}$. Any
rate pair
$$(R_1,R_2)\in \Rs_{\rm IC}(\pi_{\rm IC-I})$$
is achievable for the interference channel with confidential messages.
\end{theorem}
\begin{IEEEproof}
We provide the proof in Sec.~\ref{sec:inIC}.
\end{IEEEproof}

To derive the achievable rate region for the IC-CM, we employ an auxiliary
random variable $U$ in the sense of Han-Kobayashi \cite{HanKobayashi81}. For a
given $U$, we consider two independent stochastic encoders, that is, the
pre-coding auxiliary random variables $V_1$ and $V_2$ will be independent for a
given $U$, as given by (\ref{eq:cICa}). To ensure information-theoretic
secrecy, the achievable rate $R_1$ includes a penalty term $I(V_1;Y_2|V_2,U)$,
which is a conditional mutual information of the receiver 2's eavesdropper
channel while assuming the receiver 2 can first decode its own information.

\subsection{Broadcast Channel with Confidential Messages}

For the broadcast channel, we focus on the class of distributions
$P(u,v_1,v_2,x,y_1,y_2)$ that factor as
\begin{equation} \label{eq:cBC}
P(u)P(v_1,v_2|u)P(x|v_1,v_2)P(y_1,y_2|x).
\end{equation}
We refer to this class as $\pi_{\rm BC}$.


\begin{theorem}\label{thm:BC-O} [outer bound for BC-CM]
Let ${\Rs}_{\rm O}(\pi_{\rm BC})$ denote the union of all $(R_1,R_2)$ satisfying
\begin{subequations} \label{eq:BC-O}
\begin{align}
R_1&\ge 0,~R_2\ge 0 \notag\\
R_1&\le \min\left[
              \begin{array}{l}
                I(V_1;Y_1|U) -I(V_1;Y_2|U), \\
                I(V_1;Y_1|V_2,U) -I(V_1;Y_2|V_2,U)
              \end{array}
            \right]  \label{eq:BC-O-a}\\
R_2&\le \min\left[
              \begin{array}{l}
                I(V_2;Y_2|U)-I(V_2;Y_1| U), \\
                I(V_2;Y_2|V_1,U) -I(V_2;Y_1|V_1,U)
              \end{array}
            \right] \label{eq:BC-O-b}
\end{align}
\end{subequations}
over all distributions $P(u,v_1,v_2,x,y_1,y_2)$ in $\pi_{\rm BC}$ and auxiliary
random variables $U$, $V_1$, and $V_2$ satisfying
\begin{align}
U&\rightarrow V_1 \rightarrow X \quad \text{and} \quad U\rightarrow V_2 \rightarrow X.
\label{eq:UVX-mar}
\end{align}
For the broadcast channel  $(\Xc, P(y_1,y_2|x),  \Yc_1\times \Yc_2)$ with confidential
messages, the capacity region
$$\Cs_{\rm BC}\subseteq \Rs_{\rm O}(\pi_{\rm BC}).$$
\end{theorem}
\begin{IEEEproof}
We provide the proof of Theorem~\ref{thm:BC-O} in Sec.~\ref{sec:outer}.
\end{IEEEproof}
\begin{remark}
Outer bounds for the BC-CM and the IC-CM have a same mutual information
expression $\Rs_{\rm O}(\cdot)$, but, they are optimized over different input
distributions $\pi_{\rm BC}$ and $\pi_{\rm IC-O}$, respectively.
\end{remark}


\begin{theorem}\label{thm:InBC} [inner bound for BC-CM]
Let ${\Rs}_{\rm BC}(\pi_{\rm BC})$ denote the union of all $(R_1,R_2)$ satisfying
\begin{subequations} \label{eq:BC-IN}
\begin{align}
R_1&\ge 0,~R_2\ge 0 \notag\\
R_1 &\le I(V_1;Y_1|U)-I(V_1;V_2|U)-I(V_1;Y_2|V_2,U)  \label{eq:BC-IN-a} \\
R_2 &\le I(V_2;Y_2|U)-I(V_1;V_2|U)-I(V_2;Y_1|V_1,U) \label{eq:BC-IN-b}
\end{align}
\end{subequations}
over all distributions  $P(u,v_1,v_2,x,y_1,y_2)$ in $\pi_{\rm BC}$. Any rate pair
$$(R_1,R_2)\in \Rs_{\rm BC}(\pi_{\rm BC})$$
is achievable for the broadcast channel with confidential messages.
\end{theorem}
\begin{IEEEproof}
We provide the proof in Sec.~\ref{sec:inBC}.
\end{IEEEproof}

We note that, for a broadcast channel, we can employ joint encoding at the
transmitter. Hence, the achievable coding scheme for the BC-CM is based on the
{\it double-binning} scheme which combines the {\it Gel'fand-Pinsker binning}
\cite{GelfandPinsker80} and the {\it random binning}. To preserve
confidentiality, the achievability bounds on $R_1$ and $R_2$ each include the
penalty term $I(V_1;V_2|U)$. Without the confidentiality constraint, Marton's
inner bound \cite{Marton79} on the broadcast channel illustrates only that the
sum rate has the penalty term $I(V_1;V_2|U)$. To ensure information-theoretic
secrecy, the proposed coding scheme pays ``double'' when jointly encoding at
the transmitter.

\begin{example}\label{ex:ls} [less noisy broadcast channel]
Consider a special class of broadcast channels in which the channel
$X\rightarrow Y_1$ is {\it less noisy} than the channel $X\rightarrow Y_2$,
i.e.,
\begin{align}
I(V;Y_1) \ge I(V;Y_2) \label{eq:def-ln}
\end{align}
for every $V \rightarrow X \rightarrow (Y_1, Y_2)$ \cite{CsiszarKorner78}. We first
consider the outer bound of the less noisy BC-CM. Based on the Markov chains in
(\ref{eq:UVX-mar}) and the definition (\ref{eq:def-ln}), we have
\begin{align*}
I(V_1; Y_1|U=u)&\ge I(V_1; Y_2|U=u)\\
I(V_2; Y_1|U=u)&\ge I(V_2; Y_2|U=u),
\end{align*}
which implies that
\begin{align*}
I(V_1; Y_1|U)&\ge I(V_1; Y_2|U)\\
I(V_2; Y_1|U)&\ge I(V_2; Y_2|U).
\end{align*}
Hence the outer bound can be rewritten as the union of all $(R_1,R_2)$ satisfying
\begin{subequations} \label{eq:BC-IN-Less}
\begin{align}
R_1 &\le \max_{P(X)} [I(X;Y_1)-I(X;Y_2)]   \label{eq:BC-IN-Less-a}\\
R_2 &=0 \label{eq:BC-IN-Less-b}
\end{align}
\end{subequations}
where (\ref{eq:BC-IN-Less-a}) follows from \cite[Theorem~3]{CsiszarKorner78}. Next, by
applying Theorem~\ref{thm:InBC} and setting $V_2=U=\text{const}$, we obtain the identical
rate region as (\ref{eq:BC-IN-Less}). This result implies that only the ``better'' user
can get the non-zero secrecy rate for the less noisy BC-CM. Note that, the single-antenna
Gaussian broadcast channel is a special case of the less noisy broadcast channel.
\end{example}


In the following, we consider a sufficient condition under which both $R_1$ and
$R_2$ can be strictly positive for the BC-CM.
\begin{corollary}
For a broadcast channel, if there exist a distribution
$P(u,v_1,v_2,x,y_1,y_2) \in \pi_{\rm BC}$ for which
\begin{subequations} \label{eq:BCcond}
\begin{align}
&          &  I(V_1;Y_1|U)&>I(V_1;Y_2,V_2|U)&\\
&\text{and}&  I(V_2;Y_2|U)&>I(V_2;Y_1,V_1|U),&
\end{align}
\end{subequations}
then both receivers can achieve strictly positive rates while ensuring
information-theoretic secrecy.
\end{corollary}
\begin{IEEEproof}
The result is obtained by applying Theorem~\ref{thm:InBC} and by
setting $R_1>0$ and $R_2>0$.
\end{IEEEproof}
More recently, motivated by this work, the multiple-antenna Gaussian
broadcast channel with confidential messages was studied in
\cite{Liu:WITS:07}. It was shown that with multiple antennas at
transmitters, strictly positive rates to both receivers can be
achieved while ensuring information-theoretic secrecy.

\subsection{Switch Channel} \label{sec:sc}

In this subsection, we obtain the secrecy capacity region for a
special case of the interference channel referred to as the switch
channel (SC). As shown in Fig.~\ref{fig:SIC}, receivers in the SC
cannot listen to both transmissions (from encoders $1$ and $2$) at
the same time. For example, each encoder may transmit at a different
frequency, while each receiver may listen only to one frequency
during each symbol time $i$. We assume that each receiver $t\in \{1,
2\}$ has a random switch $s_t\in \{1,2\}$, which chooses between $t$
and $\bar{t}$ independently at each symbol time $i$ with
probabilities
\begin{align*}
P(S_{t,i}=t)&=\tau_t \\
P(S_{t,i}=\bar{t})&=1-\tau_t, \quad i=1,\dots,n
\end{align*}
where $\bar{t}$ is the complement of $t$. Therefore, receiver $t$
listens to its own information $x_{t,i}$ from encoder $t$ whenever
$S_{t,i}=t$, while it eavesdrops the signal $x_{\bar{t},i}$ when
$S_{t,i}=\bar{t}$. By assuming that the switch state information is
available at the receiver, we have that
\begin{align}
P(y_{t,i}|x_{1,i},x_{2,i},s_{t,i})&=P(y_{t,i}|x_{1,i}){\bf
1}(s_{t,i}=1)\notag\\
&\qquad +P(y_{t,i}|x_{2,i}){\bf 1}(s_{t,i}=2)\notag\\
&=P(y_{t,i}|x_{s_{t,i},i}) \label{eq:sw-model}
\end{align}
where ${\bf 1}(\cdot)$ is the indicator function.
\begin{figure}
 \centerline{\includegraphics[width=0.7\linewidth,draft=false]{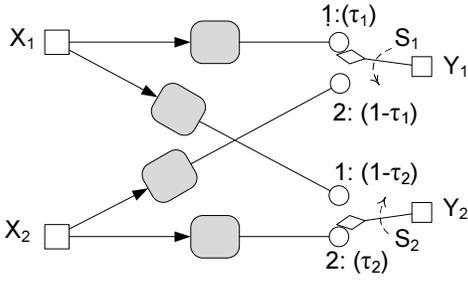}}
  \caption{Switch channel model}
  \label{fig:SIC}
\end{figure}

The switch state information $\{S_{t,i}\}^{n}_{i=1}$ is an i.i.d.
process known at receiver $t$. Hence, we can consider $s_{t,i}$ as a
part of the channel output, i.e., we set
\begin{align}
y_{t,i}\triangleq \{z_{t,i},\,s_{t,i}\} \label{eq:swout}
\end{align}
where $z_{t,i}$ represents the received signal value at receiver
$t$. Under this setting, we have the following theorem on the
secrecy capacity region $\Cs_{\rm SC}$ of SC-CM.


\begin{theorem} \label{thm:SC}
For the switch channel with confidential messages, the capacity region
$\Cs_{\rm SC}$ is the union of all $(R_1,R_2)$ satisfying
\begin{subequations} \label{eq:BC-C}
\begin{align}
0 &\le R_1 \le   I(V_1;Y_1|U)-I(V_1;Y_2|V_2,U)  \label{eq:SC-C-a} \\
0 &\le R_2 \le   I(V_2;Y_2|U)-I(V_2;Y_1|V_1,U)\label{eq:SC-C-b}
\end{align}
\end{subequations}
over all distributions $P(u,v_1,v_2,x_1,x_2,y_1,y_2)$ in $\pi_{\rm IC-I}$.
\end{theorem}
\begin{IEEEproof}
We provide the proof in the Appendix.
\end{IEEEproof}

\begin{remark}
In SC-CM, receiver $t$ listens to the desired information during
time fraction $\tau_t$, and intercepts the other message during the
time fraction $(1-\tau_t)$. When $\tau_1=\tau_2=1$, both receivers
only listen to their own messages and thus SC-CM reduces to two
independent parallel channels without the secrecy constraints. When
$\tau_1=1$ and $\tau_2=0$, receiver $2$ acts as an eavesdropper only
and both $Y_1$ and $Y_2$ are independent with respect to the message
$W_2$. Hence, in this case, SC-CM reduces to the wiretap channel
\cite{Wyner75}.
\end{remark}

\begin{example} \label{ex:sw} [noiseless memoryless switch channel]
We assume that the channel is discrete memoryless and that the input-output
relationship at each time instant satisfies
\begin{align}
Y_{t,i}=\left\{
          \begin{array}{ll}
            X_{1,i}, & S_{t,i}=1 \\
            X_{2,i}, & S_{t,i}=2
          \end{array}
        \right. \qquad \text{for} \quad i=1,\dots,n
\end{align}
where $P(S_{t,i}=t)=\tau_t$ and $\tau_1+\tau_2\ge 1$. Theorem~\ref{thm:SC}
implies that the secrecy capacity region of this channel is:
\begin{align}
\left\{ (R_1,R_2):
          \begin{array}{l}
R_1\le (\tau_1+\tau_2-1) H(X_{1})\\
R_2\le (\tau_1+\tau_2-1) H(X_{2})
          \end{array}
        \right\}.
\end{align}
We note that here $\tau_1+\tau_2-1$ equals $\tau_1-(1-\tau_2)$, the time that
user 1 sends without user 2 listening and also equals $\tau_2-(1-\tau_1)$, the
time that user 2 sends without user 1 listening.
\end{example}

\subsection{Gaussian Interference Channel with Confidential Messages} \label{sec:gic}

We next consider a Gaussian interference channel (GIC) with
confidential messages (GIC-CM) where each node employs a single
antenna as shown in Fig.~\ref{fig:GIC-model}. We have proposed this
problem originally in \cite{Liu:ITA:07}.
\begin{figure}
 \centerline{\includegraphics[width=0.65\linewidth,draft=false]{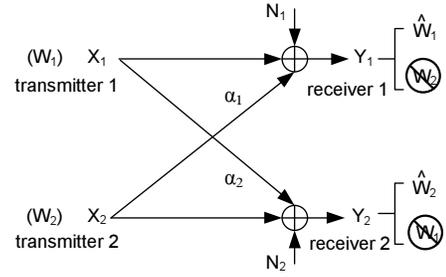}}
  \caption{Gaussian interference channel with confidential messages}
  \label{fig:GIC-model}
\end{figure}

We assume the channel input and output symbols to be from an alphabet of real
numbers. Following the standard form GIC \cite{Carleial78}, the received
symbols are
\begin{subequations} \label{eq:GICm}
\begin{align}
Y_1&= X_1+ \alpha_1 X_2 +N_1 \label{eq:GICm-a}\\
Y_2&= \alpha_2 X_1+ X_2 +N_2 \label{eq:GICm-b}
\end{align}
\end{subequations}
where $\alpha_1$ and $\alpha_2$ are normalized crossover {\it channel gains},
$X_1$ and $X_2$ are transmitted symbols from encoders $1$ and $2$ with the
average power constraint
\[\sum_{i=1}^{n}\frac{\E[X^2_{t,i}]}{n}\le P_t,\quad \text{for}~t=1,2,\]
and $N_1$ and $N_{2}$ correspond to two independent, zero-mean, unit-variance,
Gaussian noise variables. In the following, we focus on the \emph{weak}
interference channel, i.e., $0\le\alpha_1^2<1$ and $0\le\alpha_2^2<1$. We
describe three transmission schemes and their achievable rate regions under the
requirement of information-theoretic secrecy.

\subsubsection{Time-Sharing} The transmission period is divided into two
non-overlapping slots with time fractions $ \rho_1$ and $\rho_2$, where
$\rho_1\ge 0$, $\rho_2\ge 0$, and $\rho_1+ \rho_2=1$. Transmitter $t$ sends
confidential message $W_t$ in slot $t$ with time fraction $ \rho_t$, $t=1,2$.
We refer to this technique as the time-sharing scheme. We note that, in each
slot, the channel reduces to a Gaussian wiretap channel \cite{CheongHellman78}.
Let $\Rs_{\rm GIC}^{\rm [T]}$ denote the set of $(R_1,R_2)$ satisfying
\begin{subequations}
\begin{align*}
0 &\le R_1\le   \frac{\rho_1}{2} \left[\log\left(1+\frac{P_1}{\rho_1}\right)-
\log\left(1+\alpha^2_2 \frac{P_1}{\rho_1}\right)\right] \notag\\
0 &\le  R_2 \le  \frac{\rho_2}{2}
\left[\log\left(1+\frac{P_2}{\rho_2}\right)-\log\left(1+\alpha^2_1
\frac{P_2}{\rho_2}\right)\right]
\end{align*}
\end{subequations}
over all time fractions $(\rho_1,\rho_2)$ pairs. Following
\cite{CheongHellman78}, we can show that any rate pair
$$(R_1,R_2)\in \Rs_{\rm GIC}^{\rm [T]}$$ is achievable for GIC-CM.

\subsubsection{Multiplexed Transmission}
In the multiplexed transmission scheme, we allow communication links to share
the same degrees of freedom. Since we require information-theoretic secrecy for
confidential messages, no partial decoding of the other transmitter's message
is allowed at a receiver. Hence, the interference results in an increase of the
noise floor. Let
$$0\le\beta_t\le1, \quad t=1,2.$$
By independently choosing
$$V_t=X_t \sim \Nc [0,\beta_t P_t],\quad t=1,2$$
and letting $U$ serve as a convex combination operator, Theorem~\ref{thm:IC-I}
implies that any rate pair
$$(R_1,R_2)\in \Rs_{\rm GIC}^{\rm [M]}$$ is achievable for GIC-CM, where
$\Rs_{\rm GIC}^{\rm [M]}$ denotes the convex hull of the set of $(R_1,R_2)$
satisfying
\begin{subequations}
\begin{align*}
R_1&\ge 0,\quad R_2\ge 0\\
R_1&\le \frac{1}{2}\log\left(1+ \frac{\beta_1 P_1}{1+\alpha^2_1 \beta_2
P_2}\right)
-\frac{1}{2}\log(1+\alpha^2_2 \beta_1 P_1) \\
R_2&\le \frac{1}{2}\log\left(1+ \frac{\beta_2 P_2}{1+\alpha^2_2 \beta_1
P_1}\right) -\frac{1}{2}\log(1+\alpha^2_1 \beta_2 P_2)
\end{align*}
\end{subequations}
over all power-control parameters $\beta_1$ and $\beta_2$.

\subsubsection{Artificial Noise}
We next describe a scheme which allows one of the transmitters (e.g.,
transmitter $2$) to generate artificial noise. This strategy involves splitting
of the transmission power of transmitter $2$ into two parts $P_{2,M}$ and
$P_{2,A}$, where
\begin{align*}
P_{2,M}&= (1-\lambda) \beta_2 P_{2},\\
P_{2,A}&= \lambda \beta_2 P_{2}, \quad \text{and} \quad 0\le\lambda\le 1,
\end{align*}
so that transmitter $2$ encodes the confidential message with power $P_{2,M}$
and generates artificial noise with power $P_{2,A}$. The artificial noise can
spoil the received signal of receiver $2$ and, hence, protect the confidential
message of transmitter $1$. In this sense, this scheme allows \emph{transmitter
cooperation} without exchanging confidential messages.
Let $U$ serve as a convex combination operator,
\begin{align}
X_1&=V_1 \quad \text{and} \quad X_2=V_2+A_2 \label{eq:an-model}
\end{align}
where $V_1$, $V_2$, and $A_2$ are independent Gaussian random
variables:
\begin{align*}
&          & V_1 &\sim \Nc [0,\beta_1 P_1],&\\
&          & V_2 &\sim \Nc [0,P_{2,M}],&\\
&\text{and}& A_2&\sim\Nc [0,P_{2,A}].&
\end{align*}
Here $A_2$ denotes the artificial noise which cannot be predicted
and subtracted by either receiver. Since
\begin{align*}
Y_1&=X_1+\alpha_1 X_2 +N_1 \\
&=V_1+ \alpha_1 (V_2+A_2)+N_1
\end{align*}
and
\begin{align*}
Y_2&=\alpha_2 X_1+X_2 +N_1 \\
&=\alpha_2 V_1+ (V_2+A_2)+N_2,
\end{align*}
we have
\begin{align*}
  I(V_1; Y_1)&=I(V_1;\, V_1+ \alpha_1(V_2+A_2)+N_1)\notag\\
  &=h( V_1+ \alpha_1(V_2+A_2)+N_1)\notag\\
  &\quad -h(\alpha_1(V_2+A_2)+N_1)\notag\\
  &=\frac{1}{2}\log\left(1+ \frac{\beta_1 P_1}{1+\alpha_1^2 \beta_2
P_2}\right)
\end{align*}
and
\begin{align*}
  I(V_1; Y_2|V_2)&=I(V_1; \,\alpha_2 V_1+ V_2+A_2+N_2|V_2)\notag\\
  &=h(\alpha_2 V_1+A_2+N_2)-h(A_2+N_2)\notag\\
  &=\frac{1}{2}\log\left(1+\frac{\alpha^2_2\beta_1P_1}{1+\lambda\beta_2
  P_2}\right).
\end{align*}
Similarly, we can calculate
\begin{align*}
  I(V_2; Y_2)&=\frac{1}{2}\log\left[1+ \frac{(1-\lambda)\beta_2 P_2}{1+\alpha^2_2
\beta_1 P_1+\lambda\beta_2 P_2}\right]
\end{align*}
and
\begin{align*}
  I(V_2; Y_1|V_1)&=\frac{1}{2}\log
  \left[1+\frac{(1-\lambda)\alpha^2_1 \beta_2 P_2}{1+\lambda\alpha^2_1 \beta_2 P_2}\right].
\end{align*}

Applying Theorem~\ref{thm:IC-I}, we can prove that any rate pair
$$(R_1,R_2)\in \Rs_{\rm GIC}^{\rm [A]}$$ is achievable for GIC-CM, where
$\Rs_{\rm GIC}^{\rm [A]}$ denotes the convex hull of the set of $(R_1,R_2)$
satisfying
\begin{subequations}
\begin{align}
0\le R_1&\le \frac{1}{2}\log\left(1+ \frac{\beta_1 P_1}{1+\alpha^2_1 \beta_2
P_2}\right)\notag\\
&\qquad-\frac{1}{2}\log\left(1+\frac{\alpha^2_2 \beta_1 P_1}{1+\lambda\beta_2 P_2}\right) \\
0\le R_2&\le \frac{1}{2}\log\left[1+ \frac{(1-\lambda)\beta_2 P_2}{1+\alpha^2_2
\beta_1 P_1+\lambda\beta_2 P_2}\right]\notag\\
&\qquad -\frac{1}{2}\log
  \left[1+\frac{(1-\lambda)\alpha^2_1 \beta_2 P_2}{1+\lambda\alpha^2_1 \beta_2 P_2}\right]
\end{align}
\end{subequations}
over all power-control parameter pair $(\beta_1,\beta_2)$ and the
power-splitting parameter $\lambda$. Furthermore, the achievable region can be
increased by reversing the roles of transmitters $1$ and $2$.

\begin{remark}
We note that secure communication in a Gaussian channel with two senders and
two receivers was also considered in \cite{Tekin:allerton:06,Tekin:ITA:07} for
the Gaussian MAC with a wiretapper (GMAC-WT). In this setting, both messages
are to be conveyed to one of the receivers and none to the other receiver.
Although the two problem formulations differ, the absence of rate splitting in
the interference channel results in that the two proposed encoding schemes have
a closer relationship than the schemes suggested for the classical MAC and
interference channels. In fact, the encoding scheme proposed in
\cite{Tekin:allerton:06,Tekin:ITA:07} for the GMAC-WT, referred to as {\it
cooperative jamming}, and our encoding scheme which creates {\it artificial
noise} in (\ref{eq:an-model}) are the same.
\end{remark}
\begin{figure*}
\noindent
\begin{minipage}[b]{0.5\linewidth}
     \centerline{\includegraphics[width=0.9\linewidth,draft=false]{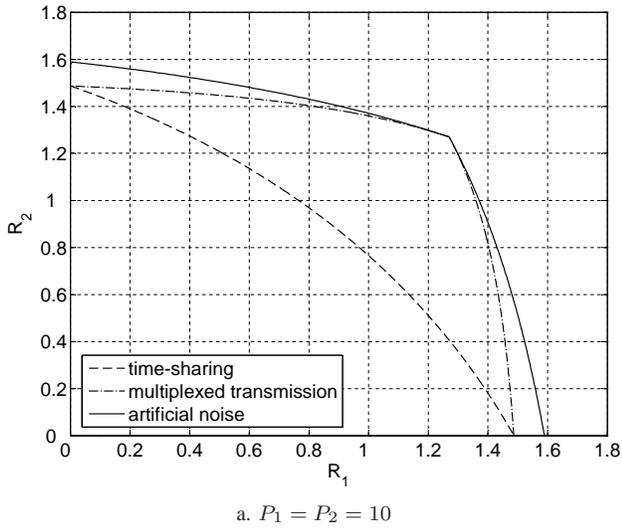}}
     \centerline{\mbox{\footnotesize a. $P_1=P_2=10$}}
  \end{minipage}\hfill
  \begin{minipage}[b]{0.5\linewidth}
     \centerline{\includegraphics[width=0.9\linewidth,draft=false]{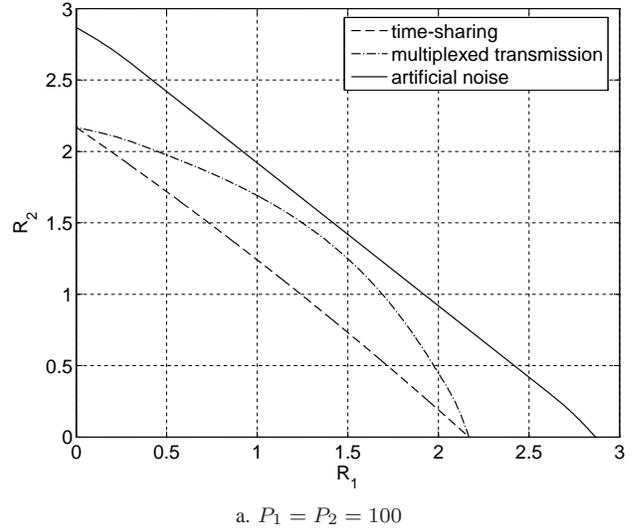}}
    \centerline{\mbox{\footnotesize a. $P_1=P_2=100$}}
  \end{minipage}
  \caption{Achievability regions for the GIC-CM ($\alpha_1=\alpha_2=0.2$).}
  \label{fig:reg-GIC}
\end{figure*}

\begin{example}
In Fig.~\ref{fig:reg-GIC}, we compare the achievable regions:
$$\Rs_{\rm GIC}^{\rm [T]},~\Rs_{\rm GIC}^{\rm [M]},~\text{and}~\Rs_{\rm GIC}^{\rm [A]}$$
by numerical calculation, for
$$P_1=P_2=10~\text{and}~\alpha_1=\alpha_2=0.2$$
in Fig.~\ref{fig:reg-GIC}.a and
$$P_1=P_2=100~\text{and}~\alpha_1=\alpha_2=0.2$$
in Fig.~\ref{fig:reg-GIC}.b. Both numerical results illustrate that the
artificial noise strategy allows for communication over larger rates, when
compared to the time-sharing and multiplexed transmission schemes.
\end{example}

\section{Outer Bound\label{sec:outer}}
In this section we prove Theorems~\ref{thm:IC-O} and~\ref{thm:BC-O}. In the
following, we derive the upper bound for $R_1$. The upper bound for $R_2$
follows by symmetry.

The basis for the outer bound derivation is the reliable transmission
requirement and the security constraint. Based on Fano's inequality
\cite{Cover}, the reliable transmission requirement (\ref{eq:MT}) implies that
\begin{subequations} \label{eq:sd}
\begin{align}
H(W_1|\Yv_1) &\le \epsilon_0 \log(M_1-1)+h(\epsilon_0) \triangleq n\delta_1.
\label{eq:sd1}\\
H(W_2|\Yv_2) &\le \epsilon_0 \log(M_2-1)+h(\epsilon_0) \triangleq n\delta_2.
\label{eq:sd2}
\end{align}
\end{subequations} where $h(x)$ is the binary entropy function. On the other
hand, the security constraint (\ref{eq:equiv1}) implies that
\begin{align}
nR_1= H(W_1) &\le H(W_1|\Yv_2)+n\epsilon_0  \label{eq:r1}.
\end{align}
In fact, the bound (\ref{eq:ob-IC}) on $R_1$ is based on the following two
different upper bounds on the equivocation $H(W_1|\Yv_2)$.

\subsection{First Bound} \label{ssec:fb}

The first upper bound is derived by applying the techniques in
\cite{CsiszarKorner78}. By using Fano's inequality (\ref{eq:sd1}), we obtain
the following bound on the equivocation
\begin{align}
H(W_1|\Yv_2) \le H(W_1|\Yv_2)-H(W_1|\Yv_1)+n\delta_1. \label{eq:cr01}
\end{align}
Let
\begin{align}
U_i=(\Yv_1^{i-1},\Ycc_2^{i+1}). \label{eq:defUI}
\end{align}
Since $(U_i,Y_{2,i})=(\Yv_1^{i-1},\Ycc_2^{i})=(U_{i-1},Y_{1,i-1})$, we have
\begin{align*}
H(W_1|U_i, Y_{2,i})-H(W_1|U_{i-1}, Y_{1,i-1})=0
\end{align*}
and we can rewrite (\ref{eq:cr01}) as follows
\begin{align}
H(W_1|\Yv_2)&\le H(W_1|\Yv_2)- H(W_1|\Yv_1) \notag\\
&\quad + \sum_{i=2}^{n} \bigl[H(W_1|U_i,
Y_{2,i})\notag\\
&\quad -H(W_1|U_{i-1}, Y_{1,i-1})\bigr] +n\delta_1. \label{eq:cr02}
\end{align}
Note that
\begin{align*}
\Yv_1&=(U_n,Y_{1,n})\quad \text{and} \quad \Yv_2=(U_1,Y_{2,1}).
\end{align*}
Hence, the bound (\ref{eq:cr02}) can be expressed as follows
\begin{align}
H(&W_1|\Yv_2)\notag\\
&\le H(W_1|U_1,Y_{2,1}) - H(W_1|U_n,Y_{1,n}) \notag\\
&\quad + \sum_{i=2}^{n}H(W_1|U_i, Y_{2,i})-
\sum_{i=1}^{n-1}H(W_1|U_{i}, Y_{1,i}) +n\delta_1\notag\\
&=\sum_{i=1}^{n} [H(W_1|U_i, Y_{2,i}) -H(W_1|U_{i},
Y_{1,i})]+n\delta_1\notag\\
&=\sum_{i=1}^{n} [I(W_1;Y_{1,i}|U_i) -I(W_1; Y_{2,i}|U_{i})]+n\delta_1\
\label{eq:cr}.
\end{align}
Inequalities (\ref{eq:r1}) and (\ref{eq:cr}) imply that
\begin{align*}
nR_1-n(\delta_1+\epsilon_0)&\le
\sum_{i=1}^{n}[I(W_1;Y_{1,i}|U_i)-I(W_1;Y_{2,i}|U_i)]. 
\end{align*}
Now, for $\delta\triangleq \delta_1+\epsilon_0$, we have
\begin{align}
R_1 &\le
\frac{1}{n}\sum_{i=1}^{n}[I(W_1;Y_{1,i}|U_i)-I(W_1;Y_{2,i}|U_i)]+\delta.\label{eq:cr1}
\end{align}
Following \cite[Chapter 14]{Cover}, we introduce a random variable $Q$
uniformly distributed over $\{1,2,\dots,n\}$ and independent of
$(W_1,W_2,\Xv_1,\Xv_2,\Yv_1,\Yv_2)$. Now we can bound $R_1$ as follows
\begin{align}
R_1&\le \frac{1}{n}\sum_{i=1}^{n}[I(W_1;Y_{1,i}|U_i,Q=i)\notag\\
&\qquad\qquad-I(W_1;Y_{2,i}|U_i,Q=i)]+\delta\notag\\
&=\sum_{i=1}^{n}P(Q=i)[I(W_1;Y_{1,Q}|U_Q,Q=i)\notag\\
&\qquad\qquad -I(W_1;Y_{2,Q}|\Yv_1^{Q-1},\Ycc_2^{Q+1},Q=i)]+\delta\notag\\
&= I(W_1;Y_{1,Q}|U_Q,Q)-I(W_1;Y_{2,Q}|U_Q,Q)+\delta. \label{eq:br2}
\end{align}
Let
\begin{align}
& U \triangleq (U_Q,Q),\quad X_1\triangleq X_{1,Q},\quad X_2\triangleq X_{2,Q},\notag\\
& Y_1\triangleq Y_{1,Q}, \quad Y_2\triangleq Y_{2,Q},\notag\\
& V_1\triangleq (W_1, U), \quad V_2\triangleq (W_2, U) . \label{eq:defv}
\end{align}
Note that, under the setting (\ref{eq:defv}), the conditional distribution of
$P(y_1,y_2|x_1,x_2)$ coincides with the original channel transition
probability. We can rewrite (\ref{eq:br2}) as
\begin{align}
R_1&\le I(V_1;Y_1|U) -I(V_1;Y_2|U)+\delta. \label{eq:outR1-1}
\end{align}

\begin{remark}
Note that we employ only  Fano's inequality (\ref{eq:sd1}) to derive the first
bound on $R_1$.
\end{remark}

\subsection{Second Bound} \label{ssec:sb}

The basic idea of the second bound can be described as follows. We assume that
a genie gives receiver~$1$ message $W_2$, while receiver $2$ attempts to
evaluate the equivocation with $W_2$ as side information.

Now, the equivocation can be upper bounded by
\begin{align}
H(W_1|\Yv_2)
&\le H(W_1,W_2|\Yv_2)\notag\\
&=   H(W_1|\Yv_2,W_2)+H(W_2|\Yv_2). \label{eq:eqv-b2}
\end{align}
By applying (\ref{eq:sd1}) and (\ref{eq:sd2}), we have
\begin{align}
H(W_1|\Yv_1)&\le n\delta_1 \quad \text{and}\quad H(W_2|\Yv_2)\le n\delta_2.
\label{eq:fa}
\end{align}
Since $H(W_1|\Yv_1,W_2)\le H(W_1|\Yv_1)$, we can further bound
(\ref{eq:eqv-b2}) as follows
\begin{align}
H(W_1|\Yv_2)
&\le   H(W_1|\Yv_2,W_2)+n\delta_2\notag\\
&\le   H(W_1|\Yv_2,W_2)-H(W_1|\Yv_1,W_2)\notag\\
&\quad +n(\delta_1+\delta_2). \label{eq:2r1-bo}
\end{align}
Let $\delta'=\delta_1+\delta_2+\epsilon_0$. Following the same approach as in
(\ref{eq:cr01})--(\ref{eq:defv}), we obtain
\begin{align}
R_1&\le I(V_1;Y_1|V_2,U) -I(V_1;Y_2|V_2,U)+\delta'. \label{eq:outR1-2}
\end{align}

\begin{remark}
In order to get the second bound on $R_1$, we employ the requirement that not
only receiver $1$ can decode the message $W_1$ successfully, but also receiver
$2$ can decode the message $W_2$ successfully in (\ref{eq:fa}) and
(\ref{eq:2r1-bo}) and, hence, we use Fano's inequalities (\ref{eq:sd1}) and
(\ref{eq:sd2}).
\end{remark}

\subsection{Outer Bound and Discussion}
Combining the two upper bounds (\ref{eq:outR1-1}) with (\ref{eq:outR1-2}) and
assuming that $\delta$ and $\delta'$ converge to $0$, we have
\begin{align}
R_1&\le \min\left[
              \begin{array}{l}
                I(V_1;Y_1|U) -I(V_1;Y_2|U), \\
                I(V_1;Y_1|V_2,U) -I(V_1;Y_2|V_2,U)
              \end{array}
            \right]. \label{eq:outR1-4}
\end{align}
Similarly, we can bound $R_2$ as
\begin{align}
R_2&\le \min\left[
              \begin{array}{l}
                I(V_2;Y_2|U)-I(V_2;Y_1| U), \\
                I(V_2;Y_2|V_1,U) -I(V_2;Y_1|V_1,U)
              \end{array}
            \right]. \label{eq:outR2}
\end{align}
Note that from (\ref{eq:defUI}) and (\ref{eq:defv}) it follows that
the joint distribution $P(u,v_1,v_2,x_1,x_2,y_1,y_2)$ factors as
(\ref{eq:cIC}) for the interference channel. For the broadcast
channel, we replace $(X_1,\,X_2)$ by $X\triangleq X_{Q}$. Now, the
joint distribution $P(u,v_1,v_2,x,y_1,y_2)$ factors as
(\ref{eq:cBC}).

To consider the sum rate we let
\begin{align*}
\Delta_1&=I(V_1;Y_1|U) -I(V_1;Y_2|U)\\
\Delta_2&=I(V_2;Y_2|U)- I(V_2;Y_1| U) \\
\Theta_1&=I(V_1;Y_1|V_2,U) -I(V_1;Y_2|V_2,U)\\
\Theta_2&=I(V_2;Y_2|V_1,U) -I(V_2;Y_1|V_1,U).
\end{align*}

The bounds (\ref{eq:outR1-4})  and (\ref{eq:outR2}) imply the the following
bounds on the sum rate:
\begin{align}
R_1+R_2&\le\Delta_1+\Delta_2,  \label{eq:sumrb1}\\
R_1+R_2&\le \Theta_1+\Theta_2 \label{eq:sumrb2}\\
R_1+R_2&\le \min[\Delta_1+\Theta_2,\,\Delta_2+\Theta_1] \label{eq:sumrb3}
\end{align}
where the bounds (\ref{eq:sumrb1}) and (\ref{eq:sumrb2}) are using either the
first bounding approach (see Sec.~\ref{ssec:fb}) or the second bounding
approach (see Sec.~\ref{ssec:sb}) only, and the bound (\ref{eq:sumrb3}) are
based on both approaches. The following lemma illustrates that the combination
sum rate bound (\ref{eq:sumrb3}) is indeed tighter than bounds
(\ref{eq:sumrb1}) and (\ref{eq:sumrb2}).
\begin{lemma} \label{lem:sumr}
\begin{align*}
\min[\Delta_1+\Theta_2,\,\Delta_2+\Theta_1] &\le \Delta_1+\Delta_2=
\Theta_1+\Theta_2.
\end{align*}
\end{lemma}
\begin{IEEEproof}
We provide the proof in the Appendix.
\end{IEEEproof}

It is interesting to further analyze the outer bound by comparing
bounds (\ref{eq:outR1-1}) and (\ref{eq:outR1-2}). By assuming that
$\delta$ and $\delta'$ converge to $0$, the difference between these
two bounds is
\begin{align}
  R_{1,\Delta}&\triangleq \Delta_1-\Theta_1\notag\\
&=I(V_1; V_2|Y_2,U)-I(V_1; V_2|Y_1,U)\notag\\
&=I(W_1; W_2|Y_2,U)-I(W_1; W_2|Y_1,U).
\end{align}
We observe that, in general, the difference between bounds (\ref{eq:outR1-1})
and (\ref{eq:outR1-2}) is non-zero.

\section{Inner Bound}  \label{sec:inner}

\subsection{Interference Channel with Confidential Messages} \label{sec:inIC}
In this subsection we derive the achievable rate region for the interference
channel. We prove that the region $\Rs_{\rm IC}(\pi_{\rm IC-I})$ is achievable.
The coding structure for the IC-CM is illustrated in Fig.~\ref{ICcodedesign}.
We employ an auxiliary random variable $U$ in the sense of Han-Kobayashi
\cite{HanKobayashi81} and two equivocation codebooks (stochastic encoders), one
for each message $W_1$ and $W_2$. Encoder $t$ maps $\vv_t$ into a channel input
$\xv_t$. More precisely, the random code generation is as follows.

Fix $P(u)$, $P(v_1|u)$ and $P(v_2|u)$, and
$$P(x_1,x_2|v_1,v_2)=P(x_1|v_1)P(x_2|v_2)$$ and let
\begin{align}
R'_1&\triangleq I(V_1;Y_2|V_2,U)-\epsilon_1 \label{eq:evr1}\\
R'_2&\triangleq I(V_2;Y_1|V_1,U)-\epsilon_1
\end{align}
where $\epsilon_1>0$ and $\epsilon_1$ is small for sufficiently large $n$.

\begin{figure}
 \centerline{\includegraphics[width=0.65\linewidth,draft=false]{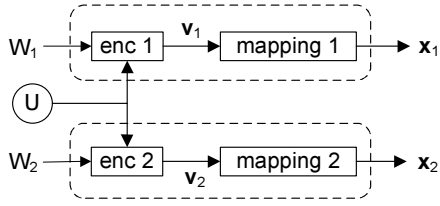}}
  \caption{Code construction for IC-CM}
  \label{ICcodedesign}
\end{figure}

\begin{itemize}
\item {[codebook generation]}
  Randomly generate a typical sequence $\uv$ with probability
  $$P(\uv)=\prod_{i=1}^{n}P(u_i),$$
  and assume that both transmitters and receivers know the time-sharing sequence $\uv$.

  For transmitter $t$, $t=1,2$,  generate $Q_t=2^{n(R_t+R'_t)}$ independent sequences $\vv_t$ each with probability
  \[P(\vv_t|\uv)=\prod_{i=1}^{n}P(v_{t,i}|u_i)\]
  and labeled as
  \begin{align}
  \vv_t(w_t,k_t), ~w_t \in \{1,\ldots , M_t \},~k_t \in \{1,\ldots , M'_t \}
  \label{eq:lab}
  \end{align}
  where $M_t=2^{nR_t}$ and $M'_t=2^{nR'_t}$. Without loss of generality, $M_t$, $M'_t$, and $Q_t$ are
  assumed to be integers.

  Let $$\Cc_t \triangleq \bigl\{\vv_t(w_t,k_t), \text{ for all } (w_t, k_t) \bigr\}$$
  be the codebook of Transmitter $t$. Its $w_t$-th sub-codebook (bin)
  $$\Cc_t(w_t) \triangleq \{\vv_t(w_t,k_t), \text{ for } k_t =1,\ldots , M'_t\bigr\}$$
  follows the partitioning in (\ref{eq:lab}).

\item {[encoding]}
  To send a message pair $$(w_1,w_2)\in \Wc_1 \times \Wc_2,$$ each transmitter employs a stochastic encoder. Encoder
  $t$ randomly chooses an element $\vv_t(w_t,k_t)$ from the sub-codebook $\Cc_t(w_t)$.
  Transmitters generate the channel input sequences based on respective mappings $P(x_1|v_1)$ and
  $P(x_2|v_2)$.

\item {[decoding]}
  Given a typical sequence $\uv$, let $\A(P_{V_t,Y_t|U})$ denote the
  set of jointly typical sequences $\vv_t$ and $\yv_t$ with respect to
  $P(v_t,y_t|u)$ \cite{Cover}. Decoder $t$ chooses $w_t$ so that
  $$(\vv_t(w_t,k_t),\yv_t)\in\A(P_{V_t,Y_t|U})$$ when such $w_t$ exists and
  and is unique; otherwise, an error is declared.
\end{itemize}

\subsubsection{Error Probability Analysis}

To bound the probability of error, we define the event
\begin{align*}
E_{t}(w_t,k_t)\triangleq \{ (\vv_t(w_t,k_t),\yv_t|\uv) \in\A(P_{V_t,Y_t|U}) \}.
\end{align*}
Without loss of generality, we can assume that transmitter $1$ sends the
message $w_1=1$ associated with the codeword $\vv_1(1,1)$, and define the
corresponding event
\begin{align*}
K_{1}\triangleq\{ \vv_1(1,1)~\text{sent}\}.
\end{align*}
The union bound on the error probability of receiver $1$ is as follows
\begin{align*}
P_{e,1}^{(n)}&\le P\left\{\bigcap_{k_1} E_{1}^{c}(1,k_1)\Big|K_{1}\right\}
+\sum_{w_1\neq 1,\,k_1}P\{E_{1}(w_1,k_1)| K_{1}\}
\notag\\
&\le P\{E_{1}^{c}(1,1)|K_{1}\}+\sum_{w_1\neq 1,\,k_1}P\{E_{1}(w_1,k_1)| K_{1}\}
\end{align*}
where $E_{1}^{c}(1,k_1)$ denotes the event
$$\{(\vv_1(1,k_1),\yv_1)  \notin \A(P_{V_1,Y_1|U})\}.$$ Following the joint asymptotic equipartition property
(AEP) \cite{Cover}, we have
\begin{align*}
P\{E_{1}^{c}(1,1)|K_{1}\}\le \epsilon,
\end{align*}
and, for $w_1\neq 1$,
\begin{align*}
P\{E_{1}(w_1,k_1)|K_{1}\}&\le2^{-n[I(V_1;Y_1|U)-\epsilon]}.
\end{align*}
Hence, we can bound the probability of error as
\begin{align*}
P_{e,1}^{(n)}& \le \epsilon+Q_1 2^{-n[I(V_1;Y_1|U)-\epsilon]} \notag\\
& =\epsilon+2^{n(R_1+R'_1)}\,2^{-n[I(V_1;Y_1|U)-\epsilon]}
\end{align*}
So, if $$R_1+R'_1< I(V_1;Y_1|U),$$ then for any $\epsilon_0>0$, $
P_{e,1}^{(n)}\le \epsilon_0$ for sufficiently large $n$. Similarly, for
receiver $2$, if $$R_2+R'_2< I(V_2;Y_2|U),$$ then $P_{e,2}^{(n)}\le \epsilon_0$
for sufficiently large $n$. Hence, $\Pe \le \epsilon_0$ as long as the rate
pair $(R_1,R_2)\in \Rs_{\rm IC}(\pi_{\rm IC-I})$.

\subsubsection{Equivocation Calculation}

To show that (\ref{eq:equiv1}) holds, we consider the following
equivocation lower bound
\begin{align}
 H(W_1|\Yv_2)&\ge  H(W_1|\Yv_2,\Vv_2,\Uv)\label{eq:ev02}
\end{align}
where inequality (\ref{eq:ev02}) is due to the fact that conditioning reduces
entropy. By applying the entropy chain rule \cite{Cover}, (\ref{eq:ev02}) can
be expanded as follows
\begin{align}
H(W_1&|\Yv_2)\notag\\
&\ge H(W_1,\Yv_2|\Vv_2,\Uv)-H(\Yv_2|\Vv_2,\Uv)\notag\\
&= H(W_1,\Vv_1,\Yv_2|\Vv_2,\Uv)\notag\\
&\quad -H(\Vv_1|\Yv_2,\Vv_2,\Uv,W_1) -H(\Yv_2|\Vv_2,\Uv)\notag\\
&=
H(W_1,\Vv_1|\Vv_2,\Uv)-H(\Vv_1|\Yv_2,\Vv_2,\Uv,W_1) \notag\\
&\quad -H(\Yv_2|\Vv_2,\Uv)+H(\Yv_2|\Vv_1,\Vv_2,\Uv,W_1). \label{eq:ev03}
\end{align}

Based on functional dependence graphs~\cite{Kramer-IT-03} and the
random code construction, we can show that the following is a Markov
chain
\begin{align*}
W_1 &\rightarrow (\Vv_1,\Vv_2,\Uv) \rightarrow \Yv_2
\end{align*}
which yields
\begin{align}
I(W_1;\Yv_2|\Vv_1,\Vv_2,\Uv)&=0. \label{eq:mk2}
\end{align}
Hence, by using (\ref{eq:ev03}) and (\ref{eq:mk2}), we obtain
\begin{align}
H(W_1|\Yv_2) &\ge
H(W_1,\Vv_1|\Vv_2,\Uv)-H(\Vv_1|\Yv_2,\Vv_2,\Uv,W_1)\notag\\
&\quad -H(\Yv_2|\Vv_2,\Uv)+H(\Yv_2|\Vv_1,\Vv_2,\Uv) \notag\\
&=H(W_1,\Vv_1|\Vv_2,\Uv)-H(\Vv_1|\Yv_2,\Vv_2,\Uv,W_1)\notag\\
&\quad -I(\Vv_1;\Yv_2|\Vv_2,\Uv)\notag\\
&\ge
H(\Vv_1|\Vv_2,\Uv)-H(\Vv_1|\Yv_2,\Vv_2,\Uv,W_1)\notag\\
&\quad -I(\Vv_1;\Yv_2|\Vv_2,\Uv).\label{eq:ev0-IB}
\end{align}

We consider the first term in (\ref{eq:ev0-IB}). Note that given $\Uv=\uv$,
$\Vv_1$ and $\Vv_2$ are independent and $\Vv_1$ has $Q_1$ possible values with
equal probability. Hence,
\begin{align}
H(\Vv_1|\Uv,\Vv_2)
&=H(\Vv_1|\Uv) \notag\\
&=\log Q_1\notag\\
&= n(R_1+R'_1). \label{eq:ev3-IB}
\end{align}

We next show that $H(\Vv_1|\Yv_2,\Vv_2,\Uv,W_1) \le n \epsilon_2$, where
$\epsilon_2$ is small for sufficiently large $n$. In order to calculate the
conditional entropy $H(\Vv_1|\Yv_2,\Vv_2,\Uv,W_1)$, we consider the following
situation. We fix $W_1=w_1$, and assume that transmitter~$1$ transmits a
codeword $\vv_1(w_1,k_1)\in \Cc_1(w_1)$, for $1\le k_1\le M'_1$, and that
receiver $2$ knows the sequences $\Vv_2=\vv_2$ and $\Uv=\uv$. Given index
$W_1=w_1$, receiver $2$ decodes the codeword $\vv_1(w_1,k_1)$ based on the
received sequence $\yv_2$. Let $\lambda(w_1)$ denote the average probability of
error of decoding the index $k_1$ at receiver $2$. Based on joint typicality
\cite[Chapter 8]{Cover}, we have the following lemma.
\begin{lemma} \label{lem:err}
$\lambda(w_1)\le \epsilon_0$ for sufficiently large $n$.
\end{lemma}
\begin{IEEEproof}
We provide the proof in the Appendix.
\end{IEEEproof}
Fano's inequality implies that
\begin{align}
\frac{1}{n}H(\Vv_1|\Yv_2,\Vv_2,\Uv,W_1=w_1)&\le \frac{1}{n}[1+\lambda(w_1)\log M'_1]\notag\\
&\le \frac{1}{n}+\epsilon_0 I(V_1;Y_2|U) \notag\\
&\triangleq \epsilon_2
\end{align}
where the second inequality follows from Lemma~\ref{lem:err} and
(\ref{eq:evr1}). Consequently,
\begin{align}
\frac{1}{n}H&(\Vv_1|\Yv_2,\Vv_2,\Uv,W_1) \notag\\
&= \frac{1}{n}\sum_{w_1\in
\Wc_1}P(W_1=w_1)H(\Vv_1|\Yv_2,\Vv_2,\Uv,W_1=w_1)\notag\\
&\le \epsilon_2. \label{eq:ev2-IB}
\end{align}

Finally, the third term in (\ref{eq:ev0-IB}) can be bounded based on the
following lemma. \noindent\begin{lemma} \label{lem:ss}
\begin{align}
I(\Vv_1;\Yv_2|\Vv_2,\Uv) \le n I(V_1;Y_2|V_2,U)+n \epsilon_3 \label{eq:vsi}
\end{align}
where $\epsilon_3$ is small for sufficiently large $n$.
\end{lemma}
\begin{IEEEproof}
We provide the proof in the Appendix.
\end{IEEEproof}

By using (\ref{eq:ev3-IB}), (\ref{eq:ev2-IB}), and (\ref{eq:vsi}),
we can rewrite (\ref{eq:ev0-IB}) as
\begin{align*}
\frac{1}{n}H(W_1|\Yv_2)\ge R_1+R'_1-I(V_1;Y_2|V_2,U)-\epsilon_2-\epsilon_3.
\end{align*}
By the definition of $R'_1$ (\ref{eq:evr1}), we have
\begin{align}
R_1-\frac{1}{n}H(W_1|\Yv_2,\Xv_2,W_2)\le \epsilon_4
\end{align}
where $\epsilon_4\triangleq \epsilon_1+\epsilon_2+\epsilon_3$, and, thus, the
security condition (\ref{eq:equiv1}) is satisfied. Following the same approach,
we can prove that (\ref{eq:equiv2}) is satisfied.

\subsection{Broadcast Channel with Confidential Messages}\label{sec:inBC}

We next prove Theorem~\ref{thm:InBC} based on the {\it double-binning} scheme
which combines the {\it Gel'fand-Pinsker binning} \cite{GelfandPinsker80} and
the {\it random binning}. In this subsection we redefine the parameters $R_1$,
$R_2$, $R'_1$, $R'_2$, $Q_1$, $Q_2$, $M_1$, and $M_2$. The coding structure for
the BC-CM is shown in Fig.~\ref{fig:BCcodedesign}.
\begin{figure}
 \centerline{\includegraphics[width=0.65\linewidth,draft=false]{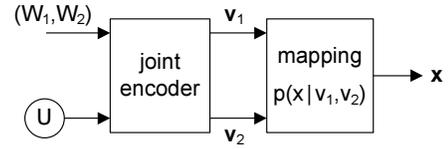}}
  \caption{Code construction for BC-CM}
  \label{fig:BCcodedesign}
\end{figure}
We employ a joint encoder to generate two equivocation codewords
$\vv_1$ and $\vv_2$, one for each message $W_1$ and $W_2$. The
equivocation codewords are mapped into the channel input $\xv$. The
details of random code generation are as follows.

We fix $P(u)$, $P(v_1|u)$ and $P(v_2|u)$, as well as $P(x|v_1,v_2)$. Let $0\le
\alpha \le 1$,
\begin{align}
R'_1&\triangleq I(V_1;Y_2|V_2,U)-\epsilon'_1 \notag\\
R'_2&\triangleq I(V_2;Y_1|V_1,U)-\epsilon'_1 \label{eq:evr3}
\end{align}
and
\begin{align}
R^{\dag} \triangleq I(V_1;V_2|U)+\epsilon'_1 \label{eq:evr2}
\end{align}
where $\epsilon'_1>0$ and $\epsilon'_1$ is small for sufficiently large $n$.

\begin{itemize}
\item {[codebook generation]}
  We generate randomly a typical sequence $\uv$ with probability $$P(\uv)=
  \prod_{i=1}^{n}P(u_i)$$
  and assume that both the transmitter and the
  receivers know the sequence $\uv$.

  We generate $Q_t=2^{n(R_t+R'_t+R^{\dag})}$ independent sequences $\vv_t$ each with probability
  \[P(\vv_t|\uv)=\prod_{i=1}^{n}P(v_{t,i}|u_i)\]
  and label them
  \begin{align}
  \vv_t(w_t,s_t,k_t),& \quad w_t \in \{1,\ldots , M_t \},~s_t \in\{1,\ldots, J_t\}, \notag\\
   &\quad \text{and} ~k_t \in \{1,\ldots , G_t
  \}. \label{eq:lab1}
  \end{align}
  where $M_t=2^{nR_t}$, $J_t=2^{nR'_t},$ and $G_t=2^{nR^{\dag}}$.
  Without loss of generality $Q_t$, $M_t$, $J_t$, and $G_t$ are considered to be integers.
\begin{figure}
  \centerline{\includegraphics[width=0.65\linewidth,draft=false]{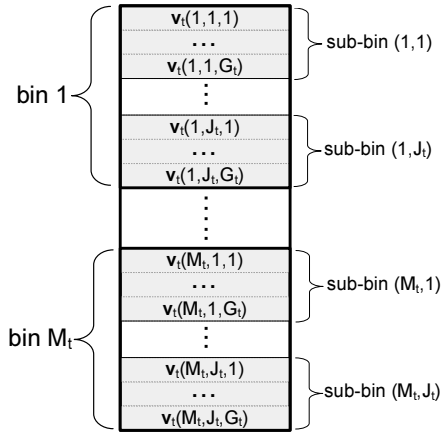}}
  \caption{Double binning}
  \label{fig:binning}
\end{figure}
  Let
  $$\Cc_t \triangleq \bigl\{\vv_t(w_t,s_t,k_t), \text{ for all } (w_t,s_t,k_t)\bigr\}$$
  denote the transmitter $t$ codebook.
  Based on the labeling in (\ref{eq:lab1}), the codebook $\Cc_t$ is
  partitioned into $M_t$ bins, and the $w_t$-th bin is
  \begin{align*}
  \Cc_t(w_t) &\triangleq \{\vv_t(w_t,s_t,k_t), \text{ for }
  s_t \in\{1,\ldots, J_t\} \notag\\
  &\qquad \text{ and } k_t \in \{1,\ldots , G_t \bigr\}.
  \end{align*}
  Furthermore, each bin $\Cc_t(w_t)$ is divided into $J_t$ sub-bins, and the $(w_t,s_t)$-th sub-bin is
  \begin{align*}
  \Cc_t(w_t,s_t) &\triangleq \bigl\{\vv_t(w_t,s_t,k_t), \text{ for }
  k_t \in \{1,\ldots , G_t\} \bigr\}.
  \end{align*}
  The double binning structure for $\vv_t$ sequences is shown in
  Fig.~\ref{fig:binning}.

\item {[encoding]}
  To send the message pair $(w_1,w_2)\in \Wc_1 \times \Wc_2 $, the transmitter employs a stochastic encoder.
  We randomly choose a sub-bin $\Cc_t(w_t,s_t)$ from the bin $\Cc_t(w_t)$, for
  $t=1,2$. Next, we select a pair $(k_1,k_2)$ so
  that
  $$\bigl(\vv_1(w_1,s_1,k_1),\vv_2(w_2,s_2,k_2)\bigr)\in \A(P_{V_1,V_2|U}),$$
  where $\A(P_{V_1,V_2|U})$ denotes, for a given typical sequence $\uv$, the
  set of jointly typical sequences $\vv_1$ and $\vv_2$ with respect to
  $P(v_1,v_2|u)$. If there are more than one such pairs, then we randomly select one.
  We generate the channel input sequence $\xv$ according to the mapping $P(x|v_1,v_2)$.

\item {[decoding]}
 For a given typical sequence $\uv$, let $\A(P_{V_t,Y_t|U})$ denote
 the set of jointly typical sequences $\vv_t$ and $\yv_t$ with respect to
 $P(v_t,y_t|u)$. Decoder $t$ chooses $w_t$ so that $(\vv_t(w_t,s_t,k_t),\yv_t)\in\A(P_{V_t,Y_t|U})$
 if such $w_t$ exists and is unique; otherwise, an error is declared.
\end{itemize}

\subsubsection{Error Probability Analysis}

Without loss of generality, we assume that the message pair is $(w_1=1, w_2=1)$
and that $s_1=s_2=1$. First, we consider the error event $T$ that the encoder
can not find an appropriate jointly typical pair, i.e.,
\begin{align*}
T &\triangleq \{\bigl(\vv_1(1,1,k_1),\vv_2(1,1,k_2)\bigr)\notin
\A(P_{V_1,V_2|U}), \notag\\
&\qquad \text{ for } s_t =1,\ldots , J_t,~k_t =1,\ldots , G_t, \text{ and }
t=1,2\}.
\end{align*}
The definition of $R^{\dag}$ in (\ref{eq:evr2}) implies that
\begin{align}
R^{\dag}>I(V_1;V_2|U).
\end{align}
Hence, following the approach of \cite{Gamal-pM}, we have that
\begin{align}
P\{T\}\le \delta_3
\end{align}
where $\delta_3>0$ and $\delta_3$ is small for sufficiently large $n$. In other
words,  the encoding is successful with probability close to $1$ as long as $n$
is large.

In the following, we assume that $(v_1(1,1,1), v_2(1,1,1))$ is sent and define
the event
\begin{align*}
K_{2}\triangleq\{(\vv_1(1,1,1), \vv_2(1,1,1))\in\A(P_{V_1,V_2|U})\}.
\end{align*}
Now, the error probability at receiver $1$ is bounded as follows
\begin{align*}
P_{e,1}^{(n)}&\le P\{T\}+(1-P\{T\})\Biggl[P\Biggl\{\bigcap_{s_1,k_1}
E_{1}^{c}(1,s_1,k_1)\Big|K_{2}\Biggr\} \notag \\
&\qquad +\sum_{w_1\neq 1}\sum_{s_1,k_1}P\{E_{1}(w_1,s_1,k_1)|
K_{2}\}\Biggr]\notag\\
&\le P\{T\}+P\{E_{1}^{c}(1,1,1)|K_{2}\}\notag \\
&\qquad +\sum_{w_1\neq 1}\sum_{s_1,k_1}P\{E_{1}(w_1,s_1,k_1)| K_{2}\}
\end{align*}
where
$$E_{t}(w_t,s_t,k_t)= \{ (\vv_t(w_t,s_t,k_t),\yv_t) \in\A(P_{V_t,Y_t|U}) \}.$$
Joint typicality \cite[Chapter 14]{Cover} implies that
\begin{align*}
P\{E_{1}^{c}(1,1,1)|K_{2}\}&\le \epsilon, \\
P\{E_{1}(w_1,s_1,k_1)|K_{2}\} &\le 2^{-n[I(V_1;Y_1|U)-\epsilon]} \quad
\text{for } w_1\neq 1.
\end{align*}
Hence, we can bound the probability of error as
\begin{align}
P_{e,1}^{(n)}& \le \delta_3+\epsilon+Q_1 2^{-n[I(V_1;Y_1|U)-\epsilon]} \notag\\
& =\delta_3+\epsilon+2^{n(R_1+R'_1+R^{\dag})}\,2^{-n[I(V_1;Y_1|U)-\epsilon]}
\end{align}
So, if
\begin{align}
R_1+R'_1+R^{\dag}< I(V_1;Y_1|U), \label{eq:cpe1BC}
\end{align}
then for any $\epsilon_0>0$, $P_{e,1}^{(n)}\le \epsilon_0$ for sufficiently
large $n$. Similarly, for receiver $2$, if
\begin{align}
R_2+R'_2+R^{\dag}< I(V_2;Y_2|U), \label{eq:cpe2BC}
\end{align}
then $P_{e,2}^{(n)}\le \epsilon_0$ for sufficiently large $n$. Hence, (\ref{eq:defPe}),
(\ref{eq:evr3}), (\ref{eq:evr2}), (\ref{eq:cpe1BC}), and (\ref{eq:cpe2BC}) imply that
$\Pe \le \epsilon_0$ as long as the rate pair $(R_1,R_2)\in \Rs_{\rm BC}(\pi_{\rm BC}).$

\subsubsection{Equivocation Calculation}

We next prove that the secrecy requirement (\ref{eq:equiv1}) holds
for BC-CM. Following the same approach as
(\ref{eq:ev02})--(\ref{eq:ev0-IB}), we have
\begin{align}
H(W_1|\Yv_2)&\ge
H(\Vv_1|\Vv_2,\Uv)-H(\Vv_1|\Yv_2,\Vv_2,\Uv,W_1)\notag \\
&\qquad -I(\Vv_1;\Yv_2|\Vv_2,\Uv)\label{eq:evBC1}.
\end{align}
Consider the first term in (\ref{eq:evBC1})
\begin{align*}
H(\Vv_1|\Uv,\Vv_2)&=H(\Vv_1|\Uv)-I(\Vv_1;\Vv_2|\Uv).
\end{align*}
Note that given $\Uv=\uv$, $\Vv_1$ attains $Q_1$ possible values with equal
probability. Hence, we have $H(\Vv_1|\Uv)=\log Q_1.$ Using the same approach as
in Lemma~\ref{lem:ss}, we can obtain
\begin{align}
I(\Vv_1;\Vv_2|\Uv) &\le n I(V_1;V_2|U)+n \epsilon'_2. \label{eq:evBC2}
\end{align}
Hence, by the definition of $R^{\dag}$ in (\ref{eq:evr2}), we have
\begin{align}
H(\Vv_1|\Uv,\Vv_2)&= \log Q_1-I(\Vv_1;\Vv_2|\Uv)\notag\\
&\ge n(R_1+R'_1+R^{\dag})-n I(V_1;V_2|U)-n \epsilon'_2\notag\\
&\ge n(R_1+R'_1-\epsilon'_2). \label{eq:evBC3}
\end{align}

Following joint typicality \cite{Cover}, (\ref{eq:ev2-IB}) implies
$$H(\Vv_1|\Yv_2,\Vv_2,\Uv,W_1)\le n\epsilon'_3$$
where $\epsilon'_3$ is small for sufficiently large $n$. Applying
Lemma~\ref{lem:ss}, the third term in (\ref{eq:evBC1}) can be bounded as
\begin{align}
I(\Vv_1;\Yv_2|\Vv_2,\Uv) &\le n I(V_1;Y_2|V_2,U)+n \epsilon'_4
\notag\\
&=n(R'_1+\epsilon'_1+\epsilon'_4)\label{eq:evBC4}
\end{align}
where $\epsilon'_4$ is small for sufficiently large $n$ and the equality
(\ref{eq:evBC4}) follows from the definition (\ref{eq:evr3}). Hence, by using
(\ref{eq:evBC2}), (\ref{eq:evBC3}), and (\ref{eq:evBC4}), we can rewrite
(\ref{eq:evBC1}) as
\begin{align*}
\frac{1}{n}H(W_1|\Yv_2)\ge R_1-\epsilon'_5
\end{align*}
where $\epsilon'_5\triangleq \epsilon'_1+\epsilon'_2+\epsilon'_3+\epsilon'_4$,
and thus the security condition (\ref{eq:equiv1}) is satisfied. Following the
same approach, we can prove that (\ref{eq:equiv2}) also holds.

\section{Conclusion} \label{sec:con}

We derived the outer and the inner bounds on the capacity of the
interference and broadcast channels with confidential messages. The
obtained results offer insights into the two communication problems.
The difference in the outer bound reflects the fact that the joint
encoding at the transmitter can only be performed in the BC-CM
whereas in the IC-CM, encoders offer independent channel inputs. The
achievability proof suggests the code construction appropriate for
these channel. We presented a special case of IC-CM for which the
two bounds meet to describe the capacity region. We proposed and
compared several transmission schemes for Gaussian interference
channels under information-theoretic secrecy. In particular, the
encoding scheme in which transmitters dedicate some of their power
to create artificial noise was shown to outperform both time-sharing
and simultaneous transmission of messages sent with the optimal
power. However, constructing practical wiretap codes that can
achieve the derived rates is a challenging problem. Code
constructions for a binary-input Gaussian wiretap channel have
recently been proposed in \cite{Liu:ITW:07}.


 \appendix


\begin{IEEEproof} {\bf (Lemma~\ref{lem:sumr})}
By the definition of $\Delta_1$, we have
\begin{align}
\Delta_1&=I(V_1;Y_1|U) -I(V_1;Y_2|U)\notag\\
&=I(V_1,V_2;Y_1|U)-I(V_2;Y_1|V_1,U) \notag \\
&\qquad -I(V_1,V_2;Y_2|U)+I(V_2;Y_2|V_1,U). \label{eq:pp1}
\end{align}
Similarly,
\begin{align}
\Delta_2&=I(V_2;Y_2|U) -I(V_2;Y_1|U)\notag\\
&=I(V_1,V_2;Y_2|U)-I(V_1;Y_2|V_2,U) \notag \\
&\qquad -I(V_1,V_2;Y_2|U)+I(V_1;Y_1|V_2,U).\label{eq:pp2}
\end{align}
(\ref{eq:pp1}) and (\ref{eq:pp2}) imply that
\begin{align}
\Delta_1+\Delta_2
&=-I(V_2;Y_1|V_1,U)+I(V_2;Y_2|V_1,U)\notag \\
&\qquad -I(V_1;Y_2|V_2,U)+I(V_1;Y_1|V_2,U)\notag\\
&=\Theta_2+\Theta_1.
\end{align}
Note that
\begin{align*}
2(\Delta_1+\Delta_2)&=2(\Theta_1+\Theta_2)\\
&=(\Delta_1+\Theta_2)+(\Delta_2+\Theta_1)
\end{align*}
Hence,
$$\min[\Delta_1+\Theta_2,\,\Delta_2+\Theta_1] \le \Delta_1+\Delta_2
=\Theta_1+\Theta_2.$$
We have the derived results.
\end{IEEEproof}

\bigskip

\begin{IEEEproof} {\bf (Lemma~\ref{lem:err})}
For a given typical sequence pair $(\vv_2,\uv)$, let $\A(P_{V_1,Y_2|V_2,U})$
denote the set of jointly typical sequences $\vv_1$ and $\yv_2$ with respect to
$P(v_1,y_2|v_2,u)$. For a given $W_1=w_1$, decoder $2$ chooses $k_1$ so that
$$(\vv_1(w_1,k_1), \yv_2)\in\A(P_{V_1,Y_2|V_2,U})$$
if such $k_1$ exists and is unique; otherwise, an error is declared. Define the
event
\begin{align*}
\hat{E}(k_1)=\{ (\vv_1(w_1,k_1),\yv_2)    \in\A(P_{V_1,Y_2|V_2,U}) \}.
\end{align*}
Without loss of generality, we assume that $\vv_1(w_1,k_1=1)$ was sent, and
define the event
\begin{align*}
\hat{K}_{1}=\bigl\{\vv_1(w_1,1)~\text{sent}\bigr\}.
\end{align*}
Hence
\begin{align*}
\lambda(w_1)&\le P\{\hat{E}^{c}(k_1=1)|\hat{K}_{1}\}+\sum_{k_1\neq 1}
P\{\hat{E}(k_1)|\hat{K}_{1}\}
\end{align*}
where $\hat{E}^{c}(k_1=1)$ denotes the event
$$\{(\vv_1(w_1,1), \yv_2)\notin \A(P_{V_1,Y_2|V_2,U})\}.$$
Following the joint AEP \cite{Cover}, we have
\begin{align*}
P\{\hat{E}^{c}(k_1=1)|\hat{K}_{1}\}\le \epsilon,
\end{align*}
and, for $k_1\neq 1$,
\begin{align*}
P\{\hat{E}(k_1)|\hat{K}_{1}\}\le 2^{-n[I(V_1;Y_2|V_2,U)-\epsilon]}.
\end{align*}
Now, we can bound the probability of error as
\begin{align*}
\lambda(w_1)& \le \epsilon+M'_1 2^{-n[I(V_1;Y_2|V_2,U)-\epsilon]} \\
& \le \epsilon+2^{nR'_1}\,2^{-n[I(V_1;Y_2|V_2,U)-\epsilon]}.
\end{align*}
Note that $R'_1=I(V_1;Y_2|V_2,U)-\epsilon_1$. Hence, by choosing
$\epsilon_1>\epsilon$, we have
\begin{align*}
\lambda(w_1)\le \epsilon_0
\end{align*}
where $\epsilon_0$ is small for sufficiently large $n$.
\end{IEEEproof}

\bigskip

\begin{IEEEproof}  {\bf (Lemma~\ref{lem:ss})} Let $\A(P_{U,V_1,V_2,Y_2})$ denote the set of typical sequences $(\uv,
\vv_1,\vv_2,\yv_2)$ with respect to $P(u,v_1,v_2,y_2)$, and
\begin{align*}
\mu(\uv, \vv_1,&\vv_2,\yv_2)=\left\{
          \begin{array}{ll}
            1, & (\uv, \vv_1,\vv_2,\yv_2) \notin \A(P_{U,V_1,V_2,Y_2}) \\
            0, & \hbox{otherwise}
          \end{array}
        \right.
\end{align*}
be the corresponding indicator function.

We expand $I(\Vv_1;\Yv_2|\Vv_2,\Uv)$ as
\begin{align}
I(\Vv_1;\Yv_2|\Vv_2,\Uv)
&\le I(\Vv_1,\mu;\Yv_2|\Vv_2,\Uv)\notag\\
&= I(\Vv_1;\Yv_2|\Vv_2,\Uv,\mu)+ I(\mu; \Yv_2|\Vv_2,\Uv)\notag\\
&=\sum_{j=0}^{1}P(\mu=j)I(\Vv_1;\Yv_2|\Vv_2,\Uv,\mu=j) \notag \\
&\qquad + I(\mu; \Yv_2|\Vv_2,\Uv) \label{eq:tm0}
\end{align}
Note that
\begin{align}
P(\mu&=1)I(\Vv_1;\Yv_2|\Vv_2,\Uv,\mu=1) \notag\\
&\le n P\bigl[(\uv,\vv_1,\vv_2,\yv_2) \notin \A(P_{U,V_1,V_2,Y_2})\bigr] \log|{\Yc}_2| \notag\\
&\le n\epsilon  \log|{\Yc}_2|\,, \label{eq:tm1}
\end{align}
and
\begin{align}
I(\mu; \Yv_2|\Vv_2,\Uv)\le H(\mu)\le 1. \label{eq:tm2}
\end{align}
We only consider the term $P(\mu=0)I(\Vv_1;\Yv_2|\Vv_2,\Uv,\mu=0)$. Following
the sequence joint typicality properties \cite{Cover}, we have
\begin{align}
P(\mu&=0)I(\Vv_1;\Yv_2|\Vv_2,\Uv,\mu=0)\notag\\
&\le I(\Vv_1;\Yv_2|\Vv_2,\Uv,\mu=0) \notag \\
&= \sum_{(\uv,\vv_1,\vv_2,\yv_2)\in\A}P(\uv,\vv_1,\vv_2,\yv_2)[\log
P(\vv_1,\yv_2|\vv_2,\uv)\notag\\
&\qquad -\log P(\yv_2|\vv_2,\uv)-\log P(\vv_1|\vv_2,\uv)] \notag\\
&\le n[H(Y_2|V_2,U)+H(V_1|V_2,U)\notag\\
&\qquad \qquad  \qquad  -H(V_1,Y_2|V_2,U) +3\epsilon] \notag \\
&=nI(V_1;Y_2|V_2,U)+3\epsilon.\label{eq:tm3}
\end{align}
Combining (\ref{eq:tm0}), (\ref{eq:tm1}), (\ref{eq:tm2}), and (\ref{eq:tm3}),
we have the desired result
\begin{align*}
I(\Vv_1;\Yv_2|\Vv_2\Uv) &\le n
I(V_1;Y_2|V_2,U)\notag\\
&\qquad +n\Bigl(\epsilon\log|{\Yc}_2|+3\epsilon+\frac{1}{n}\Bigr)\\
&=n I(V_1;Y_2|V_2,U)+n\epsilon_3
\end{align*}
where
\begin{align*}
\epsilon_3 &\triangleq \epsilon\log|{\Yc}_2|+3\epsilon+\frac{1}{n}.
\end{align*}
\end{IEEEproof}

\begin{IEEEproof} {\bf (Theorem~\ref{thm:SC})}
Since the switch channel is a special case of the interference channel, we
focus on the outer bound (\ref{eq:ob-IC}) and the inner bound (\ref{eq:IC-IN})
and prove that
$$\Rs_{\rm O}(\pi_{\rm IC-O})=\Rs_{\rm IC}(\pi_{\rm IC-I})$$
for the SC-CM case.

We note that the distribution $\pi_{\rm IC-I}$ implies that, for a
given $U$, auxiliary random variables $V_1$ and $V_2$ are
independent, but this may not hold for the distribution $\pi_{\rm
IC-O}$. Hence, we need to first show that the condition
\begin{align}
I(V_1;V_2|U)=0 \label{eq:sw-cd1}
\end{align}
holds in the outer bound for SC-CM. Furthermore, if
\begin{align}
I(V_1;V_2|Y_2,U)=0 \label{eq:sw-cd2}
\end{align}
also holds in the outer bound for SC-CM, then we have
\begin{align}
I(V_1;Y_2|V_2,U)&=I(V_1;Y_2|U)+I(V_1;V_2|Y_2,U)\notag\\
&\qquad -I(V_1;V_2|U) \notag \\
&=I(V_1;Y_2|U),\notag\\
I(V_2;Y_2|V_1,U)&=I(V_2;Y_2|U)+I(V_1;V_2|Y_2,U)\notag\\
&\qquad -I(V_1;V_2|U) \notag \\
&=I(V_2;Y_2|U),
\end{align}
that is, the outer bound (\ref{eq:ob-IC}) meets the inner bound
(\ref{eq:IC-IN}).

Now, we prove that conditions (\ref{eq:sw-cd1}) and (\ref{eq:sw-cd2}) holds in
the outer bound for SC-CM. By definitions (\ref{eq:defUI}) and (\ref{eq:defv}),
we need to show that
\begin{align}
I(W_1;W_2|U_i)&=0 \label{eq:scc1}\\
I(W_1;W_2|U_i,Y_{2,i})&=0 \label{eq:scc2}
\end{align}
where $U_i=\{\Yv_1^{i-1},\Ycc_2^{i+1}\}$. We first prove the equality
(\ref{eq:scc1}). Following the switch output definition (\ref{eq:swout}), we
have
\begin{align}
\{\Yv_1^{i-1},\Ycc_2^{i+1}\}=\{\Zv_1^{i-1},\Zcc_2^{i+1},\Sv_1^{i-1},\Scc_2^{i+1}\}
\end{align}
and hence,
\begin{align}
I(W_1&;W_2|U_i)\\
&=I(W_1;W_2|\Zv_1^{i-1},\Zcc_2^{i+1},\Sv_1^{i-1},\Scc_2^{i+1})\notag\\
&=\sum_{\sv_1^{i-1}}\sum_{\scc_2^{i+1}}P(\Sv_1^{i-1}=\sv_1^{i-1},\,\Scc_2^{i+1}=\scc_2^{i+1})
\notag\\
&\qquad I(W_1;W_2|\Zv_1^{i-1},\Zcc_2^{i+1},\sv_1^{i-1},\scc_2^{i+1})\notag\\
&=\sum_{\sv_1^{i-1}}\sum_{\scc_2^{i+1}}\Biggl[\prod_{j=1}^{i-1}
P(S_{1,j}=s_{1,j})\prod_{k=i+1}^{n}P(S_{2,k}=s_{2,k})\Biggr]\notag\\
&\qquad I(W_1;W_2|\Zv_1^{i-1},\Zcc_2^{i+1},\sv_1^{i-1},\scc_2^{i+1}).
\label{eq:swid}
\end{align}
Now, for a given $s_{t,i}$, the switch channel model (\ref{eq:sw-model})
implies that $z_{t,i}$ only depend on the channel input $x_{s_{t,i},i}$. By
using functional dependence graphs~\cite{Kramer-IT-03}, we can easily verify
that
\[I(W_1;W_2|\Zv_1^{i-1},\Zcc_2^{i+1},\sv_1^{i-1},\scc_2^{i+1})=0\]
for fixed switch state information $\sv_1^{i-1}$ and $\scc_2^{i+1}$.
Hence, (\ref{eq:swid}) implies that $I(W_1;W_2|U_i)=0$. Following
the same approach, we can prove the equality (\ref{eq:scc2}).
Therefore, we have the desired result.
\end{IEEEproof}

\section*{ACKNOWLEDGMENT}

The authors would like to thank Professor Shlomo Shamai (Shitz) of
the Technion, Gerhard Kramer, Bell Labs, Alcatel-Lucent, and Chandra
Nair, Chinese University of Hong Kong for their useful comments
about the proof of the outer bound.

\bibliographystyle{IEEEtran}
\bibliography{refICCM}

\end{document}